\documentclass[pdfa,a4paper,UKenglish,cleveref,autoref,thm-restate]{lipics-v2021}
\usepackage{listingsutf8}
\nolinenumbers
\hideLIPIcs

\lstdefinelanguage{Lean4}{
keywords={inductive, structure, where, Type, example, fun, theorem, noncomputable, def, extends, class, instance},
keywordstyle=\color{blue}\bfseries,
identifierstyle=\color{black},
commentstyle=\color{gray},
stringstyle=\color{red},
morecomment=[l][\color{gray}]{--},
morecomment=[s][\color{gray}]{/-}{-/}
}
\title{Canonical for Automated Theorem Proving in Lean}

\author{Chase Norman}{Carnegie Mellon University, Pittsburgh, PA, USA \and \url{http://chasenorman.com}}{chasen@cmu.edu}{https://orcid.org/0000-0001-8954-3770}{}
\author{Jeremy Avigad}{Carnegie Mellon University, Pittsburgh, PA, USA \and \url{https://www.andrew.cmu.edu/user/avigad/}}{avigad@cmu.edu}{https://orcid.org/0000-0003-1275-315X}{}

\authorrunning{C. Norman and J. Avigad}
\Copyright{Chase Norman and Jeremy Avigad}



\supplementdetails[cite={dagstuhl-artifact-24703},swhid={swh:1:dir:06a599989bad108dc52bfdff704d356badfa63d4;origin=https://github.com/chasenorman/Canonical;visit=swh:1:snp:b14b45164fd97819d6c472ac9195bf01ab92047c;anchor=swh:1:rev:d5c7fd0d8d0f1a2e4d8eed817d534725ea8e1156}]{Software}{https://github.com/chasenorman/Canonical}

\supplementdetails[cite={dagstuhl-artifact-24708},swhid={swh:1:dir:98f9920f9b96cabe1c6188b23fd3823d0ad9fb0f;origin=https://github.com/chasenorman/Canonicallean;visit=swh:1:snp:944ef1c9397eb4eb6dc2e6e7d500f38c3033a5ea;anchor=swh:1:rev:65797c0994b252ecca26934afe92766147fedbd3}]{Software}{https://github.com/chasenorman/CanonicalLean}

\acknowledgements{Canonical was developed with the direct support of Tesla Zhang, Cheng Zhang, Anna Zhang, Linxuan Ma, Asher Kornfeld, Abdalrhman Mohamed, Yue Yao, and Shubham Barghava.}

\funding{This material is based upon work supported by the National Science Foundation Graduate Research Fellowship Program under Grant No. DGE2140739. Any opinions, findings, and conclusions or recommendations expressed in this material are those of the author(s) and do not necessarily reflect the views of the National Science Foundation.}

\EventEditors{Yannick Forster and Chantal Keller}
\EventNoEds{2}
\EventLongTitle{16th International Conference on Interactive Theorem Proving (ITP 2025)}
\EventShortTitle{ITP 2025}
\EventAcronym{ITP}
\EventYear{2025}
\EventDate{September 28 -- October 1, 2025}
\EventLocation{Reykjavik, Iceland}
\EventLogo{}
\SeriesVolume{352}
\ArticleNo{14}

\begin{document}

\maketitle

\begin{abstract}
Canonical is a solver for type inhabitation in dependent type theory, that is, the problem of producing a term of a given type. We present a Lean tactic which invokes Canonical to generate proof terms and synthesize programs. The tactic supports higher-order and dependently-typed goals, structural recursion over indexed inductive types, and definitional equality. Canonical finds proofs for 84\% of Natural Number Game problems in 51 seconds total.
\end{abstract}

\section{Introduction}
\label{sec:introduction}

Interactive theorem provers (ITPs) like Lean \cite{lean}, Rocq \cite{rocq}, and Agda \cite{agda} are often built on the foundation of dependent type theory (DTT). In these systems, mathematical structures, function signatures, and theorem statements are represented as types. Therefore, a procedure to find terms of a given type in DTT can be directly used for witness/counterexample generation, program synthesis, and automated theorem proving (ATP). 

The problem of finding terms of a given type is known as \emph{type inhabitation}. Since DTT can be used to encode arbitrary mathematical theorems, determining whether a type is inhabited is undecidable. However, this generality indicates the potential of a sufficiently powerful type inhabitation solver in accelerating and simplifying formalization. 

Automation in ITPs is accomplished with metaprograms, called \emph{tactics}. Generally, a tactic is written to solve goals of a common, but specific, form that projects onto a decidable and well-understood theory. The tactic applies efficient and specialized reasoning to solve these goals, and builds a proof term which is not meant to be read by the user. 

A powerful class of automation tools are those which translate a goal into first order logic (FOL) or higher order logic (HOL), invoke a solver like Vampire \cite{vampire}, and reconstruct the proof to close the goal. These are referred to as \emph{hammers} \cite{sledgehammer, CoqHammer, holyhammer, mizarhammer, metamath_hammer}. By design, these systems can only use the built-in logical connectives, cannot directly reason with dependent types, and cannot synthesize functions or objects constructively. Hammers do not present an axiomatic proof term to the user, as such a term would be much too large to be read. 

In this paper, we introduce Canonical, the first solver for type inhabitation in dependent type theory. Canonical searches exhaustively for terms of any type involving dependent products ($\Pi$-types), structure types, inductive types, and let definitions. The algorithm, discussed in Section \ref{sec:canonical}, follows similarly to a theoretical account of sound and complete type inhabitation in DTT given by Dowek \cite{cube}, proceeding via iterative refinement. In Section \ref{sec:tactic}, we introduce a Lean tactic, \texttt{canonical}, which uses Canonical to find a proof term for the main goal. When it finds a proof term, it presents it to the user as a suggested replacement. While our implementation is not yet fully general,\footnote{Specifically, we do not translate quotient types or universe levels; see Section~\ref{sec:tactic}.} our goal is to approach soundness and completeness for the entirety of the Lean language, including definitional equality. 

In Section \ref{sec:results}, we demonstrate the performance of our tactic on the Natural Number Game for Lean 4 \cite{NNG}. Canonical achieves an 84\% on this benchmark, outperforming all existing Lean tactics. Canonical accomplishes this without any native support for natural numbers, propositional equality, logical connectives, or any other Lean definitions, and is instead provided the direct DTT representation at runtime. As such, these capabilities extend to user-defined datatypes and axiom schemas. We specifically highlight the higher-order reasoning and program synthesis capabilities of Canonical and show examples involving indexed inductive types and generalized induction. 

\section{Examples}

With each of the following example Lean statements we include a comment showing the invocation to Canonical that generated the proof term. The numeric argument is an optional timeout parameter, in seconds; we include it here to indicate a bound on the time that sufficed to generate the proof. 

\label{sec:Examples}
\paragraph*{Propositional Logic}
\begin{lstlisting}[language=Lean4,frame=single, backgroundcolor = \color{white}]
example : (A ∧ ¬A) → B :=
  -- by canonical 1
  fun a ↦ False.rec (fun t ↦ B) (a.right a.left)
\end{lstlisting}
Like other specialized tactics, Canonical is capable of reasoning with propositional connectives. The proof demonstrates how to use the eliminator for \texttt{False} as the principle of explosion, as well as how to project out of the \texttt{And} type. Canonical searches for proofs at increasing depth levels, so it typically finds a simple proof. Unless explicitly provided with double negation elimination, the proof will be constructive.

\paragraph*{Cantor's Theorem}
\begin{lstlisting}[language=Lean4,frame=single, backgroundcolor = \color{white}]
theorem Cantor (f : α → Set α) : ¬Surjective f :=
  -- by canonical 1 [false_of_a_eq_not_a, congrFun]
  fun a ↦ Exists.rec (motive := fun t ↦ False) 
    (fun w h ↦ false_of_a_eq_not_a (congrFun h w))
    (a fun a ↦ f a a → False) -- { a | a ∉ f a }
\end{lstlisting}
Cantor's theorem states that all functions from $\alpha$ to the power set of $\alpha$ are not surjective. Canonical generates a proof of this theorem given \texttt{false\_of\_a\_eq\_not\_a}, a Mathlib \cite{mathlib} lemma stating that a proposition cannot be equal to its negation, and \texttt{congrFun}, a standard library lemma stating that two equal functions are equal on any input. The generated proof follows the textbook proof, including the construction of the Cantor diagonal set in the last~line. 

\paragraph*{Type-Directed Program Synthesis}
\begin{lstlisting}[language=Lean4,frame=single, backgroundcolor = \color{white}]
inductive Vec (α : Type) : Nat → Type u where
| vnil  : Vec α 0
| vcons : α → {n : Nat} → Vec α n → Vec α (n + 1)

noncomputable def append : Vec α n → Vec α m → Vec α (m + n) :=
  -- by canonical 1
  fun a a_1 ↦ Vec.rec (motive := fun a t ↦ Vec α (m.add a)) 
    a_1 (fun a {n} a_2 a_ih ↦ Vec.vcons a a_ih) a
\end{lstlisting}

Canonical is not limited to proving propositions. In this example we have an indexed inductive type \texttt{Vec $\alpha$ n}, representing vectors of length \texttt{n} with elements of type $\alpha$. Canonical is able to synthesize the \texttt{append} function using only the type signature, due to the enforced length constraints. The definition must be marked ``noncomputable'' in Lean because the code generator presently cannot compile recursors. If the return type \texttt{Vec }$\alpha$\texttt{ (m + n)} is changed to \texttt{Vec }$\alpha$\texttt{ (n + m)}, Canonical will instead generate the function that appends the vectors in the reverse order. 

\paragraph*{Foundations}

\begin{lstlisting}[language=Lean4,frame=single, backgroundcolor = \color{white}]
theorem Eq.trans {a b c : α} (h₁ : Eq a b) (h₂ : Eq b c) : Eq a c :=
  -- by canonical 1
  Eq.rec (motive := fun a_1 t ↦ a = a_1) h₁ h₂
\end{lstlisting}

Canonical operates directly on the DTT representation at the foundations of Lean, including for essential notions like equality. So, Canonical can prove theorems like the transitivity of equality from first principles. Canonical has no special treatment of equality or any other constant in Lean. 

\paragraph*{Generalized Induction}

\begin{lstlisting}[language=Lean4,frame=single, backgroundcolor = \color{white}]
theorem List.recGen (motive : List α → List α → Prop) : ⋯ :=
  List.rec (motive := fun x ↦ ∀ y, motive x y)

theorem reverse_reverse (as : List α) : as.reverse.reverse = as :=
  -- by canonical 1 +synth [List.recGen]
  List.recGen 
    (fun a a_1 ↦ (a.reverseAux a_1).reverseAux [] = a_1.reverseAux a)
    (fun y ↦ Eq.refl (y.reverseAux [])) 
    (fun head tail a y ↦ a (head :: y)) as []
\end{lstlisting}

Canonical's induction ability is driven not by dedicated support, but rather by a general inhabitation procedure for DTT. It can therefore extend to more complex induction schemes like generalized induction, where the inductive hypothesis is strengthened to universally quantify over a term. In this example, we provide Canonical with the generalized induction principle for lists, and it proves that list reversal is an involution. Note that Canonical is intelligent enough to recognize that this proof must involve reasoning about the helper function \texttt{reverseAux} rather than about \texttt{reverse} directly. The \texttt{synth} flag, discussed in Section \ref{subsec:equations}, is necessary for Canonical to use an induction predicate that is only equal to the goal type under iota reduction.

\paragraph*{Higher-Order Logic}

\begin{lstlisting}[language=Lean4,frame=single, backgroundcolor = \color{white}]
theorem sSup_inter_le [CompleteLattice α] {s t : Set α} : 
  sSup (s ∩ t) ≤ sSup s ⊓ sSup t :=
  -- by canonical 20
  sSup_le (fun a ↦ s a ∧ t a) 
    (Lattice.inf (sSup fun a ↦ s a) (sSup fun a ↦ t a))
    fun b a ↦ Lattice.le_inf b (sSup fun a ↦ s a) (sSup fun a ↦ t a)
      (le_sSup (fun a ↦ s a) b (And.left a))
      (le_sSup (fun a ↦ t a) b (And.right a))
\end{lstlisting}

Beyond generalizations of induction, Canonical is able to synthesize functions to invoke arbitrary higher-order premises. In this example we prove that for any two sets $s$ and $t$ of elements from a complete lattice, the supremum of $(s \cap t)$ is less than or equal to the minimum between the supremum of $s$ and the supremum of $t$. \texttt{sSup} is a higher order function, and \texttt{sSup\_le} and \texttt{le\_sSup} are higher order axioms of complete lattices.  

\paragraph*{Multiple Inhabitants}

\begin{lstlisting}[language=Lean4,frame=single, backgroundcolor = \color{white}]
class Group' (α : Type u) extends Semigroup α, Inv α where
  one : α
  one_mul : ∀ a : α, one * a = a
  inv_mul_cancel : ∀ a : α, a⁻¹ * a = one

example [Group' α] : MulHom α α :=
  -- by canonical 1 (count := 5)
  ⟨fun a ↦ a, fun x y ↦ Eq.refl (Mul.mul x y)⟩
  ⋯
  ⟨fun a ↦ Group'.one, fun x y ↦ Eq.rec 
    (motive := fun a t ↦ a = Mul.mul Group'.one Group'.one)
    (Eq.refl (Mul.mul Group'.one Group'.one)) 
    (Group'.one_mul Group'.one)⟩
\end{lstlisting}

Since Canonical searches over inhabitants of a goal type, it can also be used to find multiple inhabitants. In this example, we use Canonical to find the two endomorphisms that exist over all groups -- the identity map and the constant function which returns the identity element. Using the \texttt{count} flag, we obtain five inhabitants with two shown above. We use a simplified group definition, \texttt{Group'}, to avoid reasoning about auxiliary notions like exponents. 

\section{Canonical}
\label{sec:canonical}

\subsection{Format} \label{format}

The first step toward creating a search procedure for DTT is to define the format for representing terms and types. The format must be expressive enough to represent the complexity of a language like Lean, while also maintaining structural properties necessary for efficient search. 

All terms in Canonical are $\beta$-normal and $\eta$-long (BNEL). This means there are no reducible expressions, such as $(\lambda ~x . ~x) ~a$, and all references to functions must be fully applied (so, a unary function $f$ is written as $\lambda ~x. ~f ~ x$). As such, all terms appear in canonical form and Canonical will only search over terms in canonical form. A term in Canonical has only one constructor, consisting of a list of parameter bindings, a list of let declarations, a head symbol, and a list of arguments:
\begin{equation} \label{term}
\lambda ~\overline{x} . ~\texttt{let} ~ \overline{y} := \overline{M}. ~f ~ \overline{A}
\end{equation}
Note that with this definition, we have direct access to the head symbol of any term. The variables $\overline{x}$ and let declarations $\overline{y}$ are in a simultaneous declaration, such that they may depend on each other in any order, including circularly.\footnote{It is the responsibility of the encoding system to provide Canonical with a signature that is type-correct in the target language, such that this circularity is not inconsistent.} The variable $f$ must be bound by a lambda binding or let definition in the term's local context. 
A type in Canonical also has only one constructor, which is a term annotated with types for each of its bound variables:
\begin{equation} \label{type}
\Pi ~\overline{x} : \overline{X}. ~\texttt{let} ~\overline{y} : \overline{Y} := \overline{M} . ~g ~\overline{B}
\end{equation}
The types $\overline{X}$ and $\overline{Y}$ again may depend on any of the bound variables $\overline{x}$ and $\overline{y}$. This system is similar to the Logical Framework \cite{LF}, although without universes. The let definitions are not essential for understanding the theory, but are used to encode constant symbols, reducible expressions, and let expressions from Lean. For bound variable $f$ of type $\Pi ~\overline{a} : \overline{Z}. ~\texttt{let} ~\overline{y'} : \overline{Y'} := \overline{M'} . ~h ~\overline{C}$, we say that the term in line \ref{term} has the type in line \ref{type} when:
\begin{equation} \label{eq}
(h ~\overline{C})[\overline{A}[\overline{M}/\overline{y}]/\overline{a}][\overline{M'}/\overline{y'}] =_\beta (g ~\overline{B})[\overline{M}/\overline{y}]
\end{equation}
and when each of the arguments $A_i$ has the correct type:
\begin{equation} \label{typing} A_i[\overline{M}/\overline{y}] : Z_i[\overline{A}[\overline{M}/\overline{y}]/\overline{a}][\overline{M'}/\overline{y'}] \end{equation}
using the standard notion of substitution and definitional reduction. Canonical accepts a type as input, and returns terms of the input type as output. For convenience, type declarations for bound variables are optional. If the type is omitted, this indicates that Canonical should not use the variable in constructing a term. The term inside a let definition may also be omitted, in which case the binding is treated as a constant symbol. To support more complex reductions, let definitions can be tagged with auxiliary reduction information. This includes the reduction rules for each recursor or the index of a field accessed by a projection function.

The most important property of this format is \emph{static arity}. From only the type signature of a variable, we can determine the number of parameters it must be supplied. This property does not hold in Lean, as the type signature can contain definitions that expand to reveal more parameters (as with \texttt{Not} or \texttt{Set}), or the number of parameters can change (such as when the polymorphic identity function \texttt{id} is specialized to a function type). 

\subsection{Refinement}

Canonical builds a term starting from a single metavariable (which can be thought of as a ``proof hole''). At each step, we pick an unassigned metavariable, select a variable from its local context to be its head symbol, and create new metavariables for the arguments to that head symbol. This operation is called a \emph{refinement}. We continue refining until all metavariables are assigned.
\begin{figure}[h]
\centering
\[\texttt{?A} \overset{\texttt{?A} \leftarrow f}{\implies} (f ~\texttt{?B} ~\texttt{?C}) \overset{\texttt{?C} \leftarrow g}{\implies} (f ~\texttt{?B} ~(g ~(\lambda ~x ~y . ~\texttt{?D})))\]
\[\overset{\texttt{?B} \leftarrow a}{\implies} (f ~a ~(g ~(\lambda ~x ~y . ~\texttt{?D}))) \overset{\texttt{?D} \leftarrow x}{\implies} (f ~a ~(g ~(\lambda ~x ~y . ~x)))\]
\caption{An example refinement sequence.}
\end{figure}

We know the number of new metavariables needed to fully apply the head symbol by the static arity property. We can also eagerly insert the correct number of lambda bindings where necessary using the static arity property, ensuring that our generated term is $\eta$-expanded. 

It is quite important that all metavariables in the application are generated at once, such that Canonical may refine them in an arbitrary order. While synthesizing the arguments sequentially would guarantee that the type of the following metavariable is fully specified (as all metavariables it depends on have been assigned), we often prefer to synthesize arguments in the reverse order. This way, the values of the earlier metavariables is constrained through unification. 

We associate with each metavariable a typing constraint. The type of the starting metavariable is constrained to be definitionally equal to the goal type. Consider the case of a metavariable \texttt{?M} with the type in line \ref{type} being refined by function symbol $f$ of type $\Pi ~\overline{a} : \overline{Z}. ~\texttt{let} ~\overline{y'} : \overline{Y'} := \overline{M'} . ~h ~\overline{C}$. Let $\overline{\texttt{?A}}$ be the newly created metavariables to apply $f$. We generate the following equational constraint to match line \ref{eq}:
\begin{equation}\label{eqmeta}(h ~\overline{C})[\overline{\texttt{?A}}[\overline{M}/\overline{y}]/\overline{a}][\overline{M'}/\overline{y'}] =_\beta (g ~\overline{B})[\overline{M}/\overline{y}]\end{equation}
and the following typing constraints for the newly created metavariables matching line \ref{typing}:
\begin{equation}\label{typingmeta}\texttt{?A}_i[\overline{M}/\overline{y}] : Z_i[\overline{\texttt{?A}}[\overline{M}/\overline{y}]/\overline{a}][\overline{M'}/\overline{y'}]\end{equation}
at which point the original typing constraint becomes inactive. If we find that one of our equational constraints is violated, we immediately backtrack the refinement decision, reactivate the original typing constraint, and continue along another path. We do not need to backtrack the choices of metavariable, only the choices of head symbol, since a given term can be obtained by any ordering of metavariable refinement. The local context of a new metavariable $\texttt{?A}_i$ consists of the local context of $\texttt{?M}$ appended with the bound variables on the right hand side of the typing constraint in line \ref{typingmeta}. 

\subsection{Explicit Substitutions}
\label{subsec:es}
How we determine whether an equational constraint has been violated is at present underspecified. In lines \ref{eqmeta} and \ref{typingmeta}, we have made frequent use of substitutions on terms containing metavariables. The expression $\texttt{?A}_i[\overline{M}/\overline{y}]$, for instance, cannot be evaluated under the standard definition of substitution, as it has not yet been determined whether $\texttt{?A}_i$ uses the variables in $\overline{y}$. Indeed, this fact may change over the course of the algorithm as $\texttt{?A}_i$ is backtracked and reassigned. Even worse, a simple equational constraint with an application:
\begin{equation}\label{bad}
(\lambda ~x . ~f ~\texttt{?X}) ~ a =_\beta b\end{equation}
cannot be reduced as it is not clear whether \texttt{?X} contains $x$. It is essential that the algorithm be able to observe that this equation is violated since $f$ and $b$ are not equal, so that we may cut the branch and backtrack. Canonical depends on the ability to determine whether an equational constraint has been violated as early as is possible, so that we only search over partial terms that are type correct.

To support this, Canonical internally uses the calculus of \emph{explicit substitutions} \cite{es1, es2, HOU_ES}. All terms and types in Canonical are de Bruijn indexed \cite{debruijn}. To determine what a de Bruijn index refers to, we accompany all terms with an explicit substitution, which maps de Bruijn indices to variables or other terms with an explicit substitution. For example, we may write the term $(\lambda ~x . ~f ~\texttt{?X}) ~ a$ in the following way:
\[\langle 0 ~1, [\langle \lambda ~1 ~\texttt{?X}, [f] \rangle, a] \rangle\]
The first component, 0 1, indicates that we should apply the first element of the explicit substitution $\langle \lambda ~1 ~\texttt{?X}, [f] \rangle$ to the subterm $\langle 1, [\langle \lambda ~1 ~\texttt{?X}, [f] \rangle, a] \rangle$, which denotes the term $a$ in this explicit substitution. To $\beta$-reduce this, we extend the explicit substitution of the function, $[f]$, with the argument(s):
\[\langle 1 ~\texttt{?X}, [\langle 1, [\langle \lambda ~1 ~\texttt{?X}, [f] \rangle, a] \rangle, f] \rangle\]
This operation can be performed in constant time with respect to the size of the terms and the number of arguments. Furthermore, $\beta$-reduction never creates new terms, only appending existing terms to existing explicit substitutions. If the first component of this term is a metavariable, $\beta$-reduction cannot continue and we say the reduction is \emph{stuck}. The let definitions from the input format encode nicely into this representation, simply extending the explicit substitution with each definition term. 

Using explicit substitutions, we can represent substitutions on metavariables and applications on partial terms. For any term in this system, we can either determine the head symbol and arguments of that term or reduction is stuck on a metavariable. If neither side of an equation is stuck, we can check if the head symbols on either side are equal. If they are not equal, we report that the equation is violated. Otherwise, we propagate the equation into equations of the subterms:
\[f ~\overline{M} =_\beta f ~\overline{N} \implies M_i =_\beta N_i ~\text{ for all } i\]
When a metavariable is refined, we reduce the equations that are stuck on that metavariable. When this refinement is backtracked, we recover the original equation. We say a metavariable has an equation if the equation is stuck on that metavariable.

\subsubsection{Equations}
\label{subsec:equations}

Canonical does not have a dedicated unification procedure or domain-specific reasoning, so equational constraints are the only guidance that the search has. It is therefore critical that we determine how best to glean information from them. We call an equation \emph{rigid} if, when it is fully reduced, it has one side which is not stuck on a metavariable. Only rigid equations can directly constrain the search, as otherwise there are two degrees of freedom. 

As mentioned previously, Canonical also contains reduction rules for let definitions, recursors, and projections. This leaves an important question as to whether the following equation should be considered violated or stuck:
\[\texttt{Nat.rec} ~a ~(\lambda n ~h. ~b) ~\texttt{?t} =_\beta a\]
While both sides have reduced to a head symbol, we cannot be sure that the major argument to the recursor \texttt{?t} will not be assigned to 0, causing the term on the left hand side to reduce to~$a$. For completeness, it is necessary to consider the equation stuck. However, if this equation is not allowed to constrain the search, then \texttt{Nat.rec} becomes a valid head symbol for almost any metavariable, as it projects into any type and does not cause equation violations. 

By default, Canonical considers such equations to be violated, which is generally the desired behavior for type-directed problems like theorem proving. However, Canonical additionally provides a \emph{program synthesis} mode where these equations are considered stuck, so it can synthesize functions that speculatively use recursors and projections even in situations where there are rigid equations. Additionally, Canonical is always forbidden from using a constructor as the head symbol of the major argument of a recursor or projection, as the resulting term would not be in canonical form. 

\subsection{Search}

Canonical uses iterative deepening to exhaustively search for terms. Rather than directly searching in increasing order of tree depth, we use a metric which we call \emph{entropy}. Entropy is high for partial terms that we expect are difficult to complete. By searching in this order, we consider more promising branches before less promising ones.

Beginning with a certain entropy threshold, we perform depth-first search (DFS) to search through all paths below the threshold. If we anticipate that the branches from a node will be large, we can add them as jobs to be completed asynchronously with fork-join parallelism. We maintain a sufficient number of jobs such that all CPU cores are utilized on independent branches of the search tree. 

We use iterative-deepening DFS so that refinement may be performed by directly mutating the assignment of a metavariable, rather than updating a persistent data structure. The cost of this is that we must retrace the steps of the prior levels each time we increase the entropy threshold. To mitigate this, we choose the threshold such that each level searches twice as many nodes as the previous. If this factor were smaller, there would be too much overhead from the redundant searching of prior levels. If it were larger, we might overshoot the number of steps necessary to find the proof term by a significant amount. This factor of~2 minimizes the product of these effects.

\subsubsection{Entropy}

We compute our custom entropy metric separately on each metavariable of a partial term. We then define the entropy of the partial term to be the product of the entropy of all of these metavariables. The metavariables of a partial term can be split into two groups: those which have already been refined and those which have yet to be refined. We treat these two groups differently from the perspective of entropy computation.

The entropy of an already refined metavariable is equal to the number of potential refinements we could have chosen for that metavariable, i.e. the branching factor at that point in the tree search. The rationale behind this is that for a branching factor $b$, we can only dedicate a $1/b$ fraction of our computational resources to each branch, making it more difficult to find a proof term. The result is that we penalize partial terms under large amounts of branching decisions, and prioritize partial terms with more forced moves that represent a larger fraction of the entire search tree. This strategy is also employed by \texttt{sauto} in Rocq \cite{sauto}. 

For the metavariables that have yet to be refined, we would like the entropy to be an estimate the number of refinements that it will take to completely solve them. Under the assumption that these metavariables are independent, the total refinement count to complete this branch will be the product of the individual refinement counts. To perform this estimate, we take advantage of the fact that our search has likely seen many similar metavariables in the past. While each metavariable may have a unique type and local context, we can \emph{bin} metavariables into groups based on properties that have a small, finite number of combinations. Canonical bins metavariables according to whether they have a rigid equation, and their type ignoring the explicit substitution, representing the type in the original input program that this metavariable derives from. Each time a recursive DFS call returns, we accumulate search information about the metavariable being refined, such as the number of refinements performed on it or its children, whether the metavariable was successfully completed, or whether all branches resulted in an equation violation. After each iteration of the iterative-deepening search, we accumulate the statistics from all of the threads into the statistics used for entropy calculation.

Armed with these statistics, we can now compute the entropy for a metavariable that has not been refined. First, we determine the probability that this metavariable will obtain a rigid equation by the time it is refined, by comparing the count of entries in its bin against that of the equivalent bin for identical metavariables but with a rigid equation. We consider metavariables with a rigid equation to have entropy $1$, as their value is essentially determined by the rigid equation. If a metavariable has probability $p$ to gain a rigid equation, we take a weighted average of 1 with weight $p$ and the average number of refinements per completed metavariable in its bin with weight $1-p$. We bound the entropy of a metavariable by 1,000 to prevent branches from becoming so penalized that they are never explored.

We similarly bin metavariables that have been refined in order to collect statistics about how often the refinement leads to a successful completion of the metavariable. We add an additional penalty to the subterm of a refined metavariable that infrequently leads to successful completions, which is removed if a completion is found. Taking the product of the entropy values of refined metavariables and metavariables that are not refined defines the total entropy for the partial term. 

\subsubsection{Metavariable Ordering}
\label{subsec:ordering}

At each node in the DFS tree, we must decide which metavariable to attempt the possible refinements on. This decision does not need to be backtracked, as any completed term can be created by refinements of the metavariables in any order. However, it is quite important that we choose a good ordering, as it can have drastic effects on the search tree. We call this the \emph{metavariable ordering} problem, from the similar ``variable ordering'' problem in constraint satisfaction problems \cite{csp}.

In general, we would like to refine metavariables that occur later in the partial term first, since the types of later metavariables are permitted to depend on the values of earlier metavariables in DTT. It is typically much easier to refine the later metavariable and determine the value of the earlier metavariable by unification than it is to search over values for the earlier metavariable and check if each allows inhabitants of the later metavariable. As an example, consider a metavariable \texttt{\frenchspacing ?n : Nat} and another $\texttt{\frenchspacing ?p : ?n} = 1000$. We would much prefer to refine \texttt{?p} to \texttt{refl} and therefore determine that \texttt{?n} must be 1000 than to iterate over natural numbers $n$ and attempt proofs of $n = 1000$ for each. It is therefore essential that our explicit substitution calculus allows us to refine metavariables in an arbitrary order, even when the metavariables in the type of another have yet to be assigned. 

Beyond this, our priority is to terminate this search branch as quickly as possible, in the fail-fast style of SAT solving \cite{SAT}. We unconditionally refine metavariables with rigid equations, as they usually either have a single refinement or no refinements at all, terminating the branch. Furthermore, this propagates the equation potentially generating constraints on other metavariables. This is the equivalent of unit propagation in SAT solving. If no metavariables have rigid equations, we look for a metavariable with a high probability of having no valid refinements, using the metavariable's statistics. 

Barring the rare case of an immediately terminating branch, the only way to stop the search is by increasing the entropy of the partial term as quickly as possible, to convince the DFS that the branch is unpromising. To do this, we prioritize metavariables with high entropy, as they likely are the most difficult to solve and harbor child metavariables with high entropy which accelerate the branch toward termination. If a metavariable has sufficiently high entropy, we prefer it over a metavariable that occurs later in the partial term but with less entropy. So, when selecting variables for refinement, we prefer selections that keep the entropy low, and when selecting the metavariable to refine, we prefer selections that increase the entropy as fast as possible. 

\section{Lean Tactic}
\label{sec:tactic}

Canonical is a Rust program, which is linked to Lean via the Foreign Function Interface. For structures in the input format of Canonical (subsection \ref{format}), like terms and types, we create equivalent structures in Lean:

\begin{lstlisting}[language=Lean4,frame=single, backgroundcolor = \color{white}]
    structure Term where            structure Typ where
        params: Array Var               params: Array (Option Typ)
        lets: Array Let                 lets: Array (Option Typ)
        head: String                    codomain: Term
        args: Array Term
\end{lstlisting}

Translating a Lean $\texttt{Expr}$ into these structures proceeds by pattern matching on the $\texttt{Expr}$:

\begin{lstlisting}[language=Lean4, frame=single, backgroundcolor = \color{white}]
inductive Expr where
  | forallE: Name → Expr → Expr → BinderInfo → Expr
  | lam: Name → Expr → Expr → BinderInfo → Expr
  | letE: Name → Expr → Expr → Expr → Bool → Expr
  | app: Expr → Expr → Expr
  | fvar: FVarId → Expr
  | const: Name → List Level → Expr
  | sort: Level → Expr
  ...
\end{lstlisting}

The \texttt{forallE} constructor represents Pi types, function types, universal quantification, or implication. When translating into a \texttt{Typ} we append \texttt{forallE} bindings to \texttt{params}, with the parameter \texttt{Typ} being the recursive translation of the type of the \texttt{forallE} binding. Likewise, \texttt{lam} represents a lambda expression. When translating into a \texttt{Term}, we append \texttt{lam} bindings to \texttt{params}. Let definitions, with the constructor \texttt{letE}, are treated similarly but are added to \texttt{lets}, and their definition is recursively translated into a \texttt{Term}. 

The \texttt{app} constructor represents the application of a function expression to an argument expression. When translating \texttt{(app} $e_1 ~e_2$\texttt{)}, we append the translation of the argument $e_2$ to \texttt{args}, and continue translation on the function expression $e_1$. Finally, once a constant symbol (\texttt{const}) or free variable (\texttt{fvar}) is reached, we set \texttt{head} to the name of the symbol, and return the complete \texttt{Term}.

There are a number of additional considerations in this translation. If we encounter a lambda expression inside the function \texttt{Expr} of an application (i.e. we are translating a $\beta$-redex), we pair the lambda binding and latest applied argument as a let expression. For example, this translates $(\lambda x. ~x) ~y$ into $\texttt{let} ~x := y. ~x$, preventing pervasive reduction while remaining in BNEL form. 

Canonical has no built-in object-theoretic notion of Sort. We add a let definition at the top level for \texttt{Sort} with no implementation and type \texttt{Sort}. This is similar to Lean's judgment $\texttt{\frenchspacing Sort u : Sort (u + 1)}$, but removing the universe level annotation. In some circumstances this leads to unsoundness by Girard's paradox, but in practice we have not seen Canonical solve a goal in this way. When we encounter a \texttt{Nat} or \texttt{String} literal, we create a let declaration to represent it, with the appropriate type. Due to the size of literals, we generally do not include a definition for the let declaration, unless it is a natural number less than or equal to 5. In this case, we provide a definition in terms of the constructors of~Nat.

\subsection{Arrow Types}
\label{subsec:arrow}
When translating a term, if we encounter a forall expression $(a : A) \to B ~a$, we convert it to $\texttt{Pi} ~A ~(\lambda a. ~B ~a)$ for the following definition of $\texttt{Pi}$:

\begin{lstlisting}[language=Lean4,frame=single, backgroundcolor = \color{white}]
structure Pi (A : Type u) (B : A → Type v) where
    f : (a : A) → B a
\end{lstlisting}

When we reach a constant symbol or free variable, we check the number of applied arguments against the arity of the symbol. Since Lean does not have static arity, we choose the minimum arity, which is the number of forall expressions in the type of the symbol. If fewer arguments were supplied than are necessary, we $\eta$-expand by adding parameters and supplying those as arguments until the symbol is fully applied. If more arguments were supplied than the arity, this implies that the return type of the function was a $\texttt{Pi}$ type. So, we convert the applications of the excess arguments into invocations of \texttt{Pi.f}. We also ensure that each argument supplied to the constant symbol or free variable has the correct arity. If an argument has fewer parameters than necessary, we add new parameters which become arguments on its head symbol. If an argument has more parameters than necessary, the argument must be a $\texttt{Pi}$ type. So, we replace the innermost arguments with an application to $\texttt{Pi.mk}$. As an example of these translations, we translate $\texttt{id (A → B) f a}$ to:
\[\texttt{Pi.f A (} \lambda \texttt{\frenchspacing a. B) (id (Pi A (} \lambda\texttt{\frenchspacing a. B)) (Pi.mk A (} \lambda\texttt{\frenchspacing a. B) (} \lambda\texttt{\frenchspacing x. f x))) a}\]

Unnecessary uses of \texttt{Pi} negatively affect the performance of Canonical. So, we pervasively unfold definitions that alias a forall expression. For example, $\texttt{Set} ~\alpha$ is unfolded to its definition $\alpha \to \texttt{Prop}$ and \texttt{Not P} is unfolded to its definition $\texttt{P} \to \texttt{False}$. In this way, they will be translated to native $\Pi$-types in Canonical where possible. We also unfold reducible definitions and typeclass instances to reduce the complexity of the generated problem.

\subsection{Definitions}

As discussed in Section \ref{format}, all constant symbols used by Canonical must be included as top-level let declarations. As such, we must add a let declaration when we encounter a constant symbol for the first time. If the constant symbol is a definition, we recursively translate the definition and include it in the let declaration. 

In general, if we provide a let declaration with a definition, it is inadvisable to allow Canonical to use the let definition in a proof. This is because it allows for redundant ways of expressing the same term, by either using the let declaration or rewriting its definition. For instance, we do not provide a type for the let declaration \texttt{let 0 := Nat.zero} as this would allow Canonical to write 0 in two distinct ways. However, we choose to provide the type of any let declaration that uses a recursor or auxiliary recursor, as these definitions are a suitably complex abstraction. If a constant symbol is only used in the definition of a let declaration, we do not provide it a type as it is not directly relevant to the problem. 

For constant symbols with special reduction behavior, like recursors, projection functions, and constructors, we annotate the associated let declaration with information about the reduction. For instance, we recursively translate the reduction rules for recursors on inductive types. Whenever we define an inductive type (resp. structure type) we automatically define the constructor(s) and recursor (resp. projection functions). 

We also automatically add all declarations in the local context of the main goal as let declarations. Users may supply additional constant symbols to be defined in square brackets. 

It is important to emphasize that the recursive translation of constant symbols and their definitions continues until primitive declarations are reached. This can violate abstraction barriers and lead to an explosion in the number of declarations if, for instance, real numbers are used. Deciding on which definitions to include is a question of relevance filtering and is beyond the scope of this work. We currently use a faithful recursive translation for completeness, but it is straightforward to limit the definitions that Canonical receives. 

\section{Results}
\label{sec:results}

It is a rite of passage for many Lean beginners to play the Natural Number Game. This Lean tutorial involves proving elementary properties of natural numbers in 79 ``levels'' over 9 ``worlds'' about topics like addition, multiplication, exponentiation, and inequalities. The game introduces a number of essential Lean tactics, with proofs involving induction, equational rewrites, and propositional logic. 

We have composed a Lean file with the 74 unique statements proven in the Natural Number Game.\footnote{Some statements are duplicated to showcase the use of specialized tactics.} The Natural Number Game uses a custom definition of natural numbers and the operations on them for pedagogical reasons. Our file instead uses the Lean standard library definition of \texttt{Nat} to allow fair comparison with tactics that have dedicated \texttt{Nat} logic, to more accurately represent real use cases, and to allow definitional reduction of some functions that the Natural Number Game defines axiomatically. It is important to note that Canonical does not have specialization toward natural numbers or equality, and indeed (automatically) receives the inductive definition of these types separately with each problem. 

The Natural Number Game was chosen as a benchmark because it has a mixture of propositional, first-order, and higher-order reasoning problems. On first-order problems, we can observe the capabilities of Canonical compared to tools dedicated towards this task. On higher-order problems, we can observe the benefits of the generality of Canonical. Furthermore, the statements in the Natural Number Game involve simple objects that are close to the foundations of Lean, and therefore admit reasonably short proof terms. In general, the Natural Number Game represents a basic level of proficiency with the Lean language. 

We have manually selected the premises required to solve each problem and have supplied them to each tool. In most cases, these are derived from the reference solutions and consist of theorems proven in earlier levels. However, we must occasionally provide theorems from the standard library, such as \texttt{Nat.le.intro} and \texttt{Nat.le.dest} which convert between the standard library definition of $\leq$ and the one used in the natural number game. Canonical additionally requires \texttt{Classical.byContradiction} for one level in which a proof by contradiction is required. A number of problems require induction where the induction predicate is an implication. To allow canonical to represent implications, as they are written in Lean as arrow types, we provide Canonical with the following definition of \texttt{If} in these levels, similar to the definition of \texttt{Pi} in Section \ref{subsec:arrow}:

\begin{lstlisting}[language=Lean4, frame=single, backgroundcolor = \color{white}]
structure If (a b : Prop) : Prop where
  h : a → b
\end{lstlisting}

We compare against Lean tactics Aesop \cite{aesop} and Duper \cite{duper}. Since Aesop can use \texttt{simp}~\cite{lean}, we manually removed the \texttt{simp} attribute from Lean standard library theorems about natural numbers until after the associated level in the game where that theorem is proven. Duper requires that the definitional reduction rules be provided as propositional equalities, so we provide these where necessary. All of our experiments are performed on a Mac Mini with M4 Pro and 24GB of RAM. We use Lean and Mathlib version 4.16.0, and Duper version 0.0.23. We use the default options for Aesop, portfolio mode for Duper, and we provide Canonical a timeout of 15 seconds. 

\begin{table}[h]
\caption{Natural Number Game problems solved by each tactic, separated by world.}
\label{fig:table}
\centering
\begin{tabular}{|c|c|c|c|c|c|c|c|c|c|c|}
\hline
& Tut & Add & Impl  & Mul & Algo & Add2 & Pow  & LEQ  & Mul2 & Total \\
\hline
Canonical & 6/6 & 5/5 & 10/10 & 9/9 & 6/7  & 6/6  & 7/10 & 7/11 & 6/10  & 62/74 (84\%) \\
\hline
Aesop     & 6/6 & 0/5 & 9/10  & 0/9 & 4/7  & 0/6  & 2/10 & 5/11 & 1/10  & 27/74 (36\%)\\
\hline
Duper     & 6/6 & 1/5 & 9/10  & 4/9 & 7/7  & 4/6  & 5/10 & 5/11 & 4/10  & 45/74 (61\%) \\
\hline 
\end{tabular}
\end{table}

Table \ref{fig:table}  shows the problems solved by each tactic, separated by world. Aesop solves 36\% of the problems, Duper solves 61\%, and Canonical solves 84\%. This difference is primarily because Aesop and Duper are not intended to perform induction, and were not able to solve any problems requiring it. The first problem that Canonical fails on is in the Algorithm world:

\begin{lstlisting}[language=Lean4, frame=single, backgroundcolor = \color{white}]
example (a b c d e f g h : Nat) : 
(d + f) + (h + (a + c)) + (g + e + b) = a + b + c + d + e + f + g + h
\end{lstlisting}

Canonical cannot use tactics, and therefore cannot solve this problem in the intended manner with \texttt{simp}. Solving this problem therefore requires too a large number of \texttt{Eq.rec} invocations. Duper, with special support for equational reasoning, can solve this problem. Requiring too many invocations of induction or generalized induction is the most common reason that Canonical did not solve a problem. Several problems also are affected by the incompleteness in our handling of iota reduction (see Section \ref{subsec:equations}), causing Canonical to be unable to solve some problems that admit a short proof.

In total, Canonical takes 51 seconds to generate the 62 proofs. Only 9 proofs take longer than a second. While solving these 9 problems, Canonical attempts an average of 23 million refinements per second, with 2.1 million per second not resulting in a violated equation. Of these, an average of 1.2 million refinements per second resulted in a partial term still below the entropy threshold, triggering a recursive DFS call. The average branching factor is 1.6, owing to the fact that most metavariables are solved by unification (rigid equations). On these 9 problems, the final iteration of iterative deepening accounts for only 17\% of solve time, suggesting a bias for proofs to be found early in an iteration. 

We computed the length of the proofs generated by each tactic by counting the number of constant symbols, free variables, bound variables, sorts, literals, and projections in the term, excluding the types of lambda and let bindings. On problems where both Duper and Canonical found a proof, the proof found by Canonical was on average 251x shorter. For Aesop, the proofs found by Canonical were on average 18x shorter. On top of this, Canonical only uses constant symbols provided to it or those needed to define the problems statement. Aesop and Duper make frequent use of custom, domain-specific constant symbols. For instance, Aesop would use lemmas about subtraction when subtraction is not a part of the Natural Number Game. We see a potential use case for Canonical in proof minimization and repair for these reasons. 

\enlargethispage{\baselineskip}
We have also tested Canonical on definitions from the Prelude of the Lean standard library. These problems are generally not in scope for other tools, as they are often not propositions and they require reasoning directly about the inductive definitions of types in advance of basic lemmas being proven. For non-propositions, we generate multiple inhabitants using the \texttt{count} option, and consider Canonical to have solved the example if one of the inhabitants is definitionally equal to the standard library implementation. We also supply proofs that constructors of an inductive type are disequal to each other, in lieu of \texttt{noConfusion}. Of the first 50 statements in the library, Canonical solves 46 (92\%). Three failures are caused by our tactic not encoding the eta rules for structures, and one is caused by our incompleteness with respect to iota reduction. A selection of terms generated by Canonical are shown below.

\begin{lstlisting}[language=Lean4, frame=single, backgroundcolor = \color{white}]
theorem eq_false_of_ne_true : 
  {b : Bool} → Not (Eq b true) → Eq b false :=
  -- by canonical 3 [Pi]
  fun {b} a ↦ Bool.rec (motive := fun t ↦ b = t → t = false) 
    (fun a ↦ Eq.refl false)
    (fun a_1 ↦ False.rec (fun t ↦ true = false) (a a_1)) 
    b (Eq.refl b)
\end{lstlisting}

\begin{lstlisting}[language=Lean4, frame=single, backgroundcolor = \color{white}]
instance Pi.instNonempty {α : Sort u} {β : α → Sort v} 
  [(a : α) → Nonempty (β a)] : Nonempty ((a : α) → β a) :=
  -- by canonical 1 [Nonempty.intro, Classical.ofNonempty]
  Nonempty.intro fun a ↦ Classical.ofNonempty
\end{lstlisting}

\begin{lstlisting}[language=Lean4, frame=single, backgroundcolor = \color{white}]
theorem decide_eq_true [inst : Decidable p] : 
  p → Eq (decide p) true :=
  -- by canonical 1
  fun a ↦ Decidable.rec (motive := fun t ↦ decide p = true)
      (fun h ↦ False.rec (fun t ↦ false = true) (h a)) 
      (fun h ↦ Eq.refl true) inst
\end{lstlisting}

\begin{lstlisting}[language=Lean4, frame=single, backgroundcolor = \color{white}]
def ite {α : Sort u} (c : Prop) [h : Decidable c] (t e : α) : α :=
  -- by canonical 1 (count := 5)
  Decidable.rec (motive := fun t ↦ α) (fun h ↦ e) (fun h ↦ t) h
\end{lstlisting}

\section{Related Work}
\label{sec:background}

There are a number of existing theorem proving systems that are based on type inhabitation. There are \texttt{sauto} \cite{sauto} for Rocq, Agsy \cite{Agsy}, Mimer \cite{mimer}, and Auto \cite{AgdaAuto} for Agda, and TWELF~\cite{twelf} for the Logical Framework \cite{LF}. An essential subroutine for these systems is \emph{unification}, in which two terms with existentially quantified variables are made equal to each other by a substitution on those variables. An algorithm for first-order unification was described by Robinson in 1965 \cite{Robinson1965Resolution}. In the higher-order setting, unification can have multiple (or infinitely many) solutions and is undecidable \cite{undecidable, Goldfarb1981}. Nonetheless, Huet provided a procedure to enumerate higher-order unifiers in 1975 \cite{huet}. To date, no theorem provers use this algorithm on account of its inefficiency, and instead elect for decidable and efficient \emph{pattern} unification~\cite{Miller1991Unification} (or extensions thereof, e.g.~\cite{AbelPientka2018Extensions}). In particular, the restriction to pattern unification ensures that the solutions to unification problems are unique. While this handles many common cases, this means that these systems cannot perform general higher-order reasoning, such as the synthesis of induction predicates. Beyond this, these systems are not designed to handle the complexity of problems we are interested in solving with Canonical, such as those involving definitional equalities and requiring several nested recursors.

Some tools offer similar functionality for specialized use cases. ELPI \cite{elpi} is an interpreter for the higher-order logic programming language $\lambda$Prolog for use in elaboration and type inference of programming languages. It uses the higher-order capabilities of $\lambda$Prolog to instantiate metavariables in some cases, but is not intended for theorem proving on its own. An extension of the program synthesis tool Herb to dependent types \cite{dttprogsynth} can write simple recursive programs using goal-directed search with pattern unification. The Theorem Proving System \cite{TPS} has native support for simple type theory as well as first-order logic, and is able to prove a wide range of theorems with this capability. 

While these systems all have type theory support, they each sacrifice completeness or generality in some way. The only account of a complete search procedure for dependent type theory was given by Dowek \cite{cube}. Our algorithm is conceptually that of Dowek, although we use a custom explicit substitution representation for performance. To our knowledge, Dowek's algorithm has not been implemented. 

The most common approach for ATP in proof assistants is called a \emph{hammer}. In these systems, a goal is monomorphized into first-order logic \cite{MengPaulson2008, translation} which is then fed to a first-order theorem prover \cite{vampire, zipperposition2, eprover} or SMT solver \cite{z3, cvc5}. If the statement holds, a proof reconstruction tool builds a proof in the target language, often by re-proving the goal with a proof-producing first-order solver \cite{metis, duper}. This approach has been successful in languages that are built on higher-order logic \cite{sledgehammer, holyhammer, mizarhammer} but has also been implemented in Rocq \cite{CoqHammer} and Metamath \cite{metamath_hammer}. These tools are adept at solving goals in the first-order fragment of the target language. There are also standalone tactics that invoke SMT solvers to solve first-order goals \cite{leansmt, smtcoq}.

An entirely different class of ATP system are those that search over tactic proofs. Aesop~\cite{aesop} performs such a search in Lean using hand-tuned heuristics. More recently, Large Language Models have been used to predict tactics \cite{STP, goedel, leancopilot, deepseek, intern}. Since Canonical can be invoked via a Lean tactic, we hope that Canonical can be beneficial to tactic prediction tools and human formalizers alike. 

\section{Future Work}
\label{sec:future}

\enlargethispage{\baselineskip}
The \texttt{canonical} tactic applies an exhaustive backward search directly on the DTT representation of a Lean goal. It does so proficiently, but many goals are not best solved in this manner. Canonical generates cut-free proofs (i.e. it does not create its own let declarations) and it has difficulty reasoning about structure types as it must first apply the correct projection function to access the field. The solution to this is a complementary system that performs forward reasoning. Forward reasoning moves are fundamentally different from backward reasoning moves in that they do not need to be backtracked (and thus do not form a search tree) and they require a level of prescience that cannot be achieved by a fail-fast approach. For this reason, we believe a separate, learning-based technique for forward reasoning is needed. We envision Canonical as being a buffer between AI systems and type theory, such that they do not need to worry about soundness or completeness, or about goals that can easily be solved via search. 

Similarly, Canonical cannot apply tactics or other domain-specific reasoning. A direct approach to add tactic support to Canonical will not succeed, as the types of metavariables for which domain-specific reasoning would be useful often contain multiple metavariables at the time they are attempted, halting most tactics. Enriching tactics such that they may be performed progressively as refinements are made is a question of programming languages theory \cite{building_in} and is beyond the scope of this work. 

With the exception of statistical information, there is little communication between the branches of the DFS search in Canonical. Established formal methods techniques like conflict clauses in SAT \cite{SAT} and tabled logic programming in TWELF \cite{tabling} do not have an immediate generalization to this setting. Finding such a generalization would help to eliminate a substantial amount of redundancy from the search space. 

Finally, while we have targeted Lean for its popularity and usability, we do not foresee any intrinsic limitation in invoking Canonical to solve goals in other languages built on DTT, including advanced theories like cubical type theory \cite{cubical} and those with modal types~\cite{modal}. This further showcases the potential of Canonical as an AI gym \cite{pantograph, leandojo}, as data can be sourced from all of these languages (or generated, using ATP) to train models that are also language-agnostic. We leave this to future work. 

\section{Conclusion}
\label{sec:conclusion}

Canonical is the first system capable of searching through terms in DTT, the foundations of Lean, Rocq, and Agda. The expressiveness of DTT and the generality of Canonical allows it to perform generalized induction on indexed inductive types, use higher-order axioms, and generate multiple constructive inhabitants of a type -- all by exhaustive enumeration at the foundational level. We believe that search is a core technology that is useful for automated proof, program synthesis, witness/counterexample generation, pedagogy, proof minimization, and data generation for AI. Beyond this, robust goal-directed reasoning is a necessary capability to reason at the human level. 

\bibliography{p014-Norman}

\end{document}